\documentclass[a4paper,fleqn,usenatbib]{mnras}

\usepackage[T1]{fontenc}
\usepackage{ae,aecompl}

\usepackage{graphicx}	
\usepackage{amsmath}	
\usepackage{amssymb}	
\usepackage{paralist}
\usepackage{upgreek}

\title[Low Frequency Radio counterpart of HESS J1731-347]{325 and 610 MHz Radio Counterparts of SNR G353.6$-$0.7  a.k.a. HESS J1731$-$347  }

\author[Nayana et al.]{Nayana A.J.$^{1}$\thanks{E-mail: nayana@ncra.tifr.res.in}, 
Chandra, Poonam$^{1}$, 
Roy, Subhashis$^{1}$, 
Green, David A.$^{2}$,
Acero, Fabio$^{3}$,
\newauthor
Lemoine-Goumard, Marianne$^{4}$,
Marcowith, Alexandre$^{5}$,
Ray, Alak K.$^{6}$ and
\newauthor
Renaud, Matthieu$^{5}$
\\
$^{1}$National centre for Radio Astrophysics,
Tata Institute of Fundamental Research, PO Box 3, Pune 411 007, India\\
$^{2}$Astrophysics Group, Cavendish Laboratory, 19 J. J. Thomson Avenue, Cambridge CB3 0HE\\
$^{3}$Laboratoire AIM, IRFU/SAp - CEA/DRF - CNRS - Universit\'e Paris Diderot, Bat. 709, CEA-Saclay, Gif-sur-Yvette Cedex, France\\
$^{4}$Centre d'Etudes Nucl\'{e}aires de Bordeaux Gradignan, Universit\'{e} Bordeaux, CNRS/IN2P3, 33175 Gradignan, France\\
$^{5}$Laboratoire Univers et Particules de Montpellier (LUPM), Universit\'{e}  de Montpellier, CNRS/IN2P3, Montpellier, France\\
$^{6}$Tata Institute of Fundamental Research, Homi Bhabha Road, Mumbai 400001, India\\
}

\date{Accepted XXX. Received YYY; in original form ZZZ}

\pubyear{2016}

\begin{document}
\label{firstpage}
\pagerange{\pageref{firstpage}--\pageref{lastpage}}
\maketitle

\begin{abstract}
HESS J1731$-$347 a.k.a. SNR G353.6$-$0.7 is one of the five known shell-type supernova remnants (SNRs) emitting in the very high energy (VHE, Energy
$>$ 0.1 TeV) $\gamma$-ray domain.  We observed this TeV SNR  with 
the Giant Metrewave Radio Telescope (GMRT) in 1390, 610 and 325 MHz bands. 
In this paper, we report the discovery of 325 and 610 MHz radio counterparts of the  SNR HESS J1731$-$347 with the GMRT.   
Various filaments of the SNR are clearly seen in the 325 and
 610 MHz bands. However, the faintest feature in the radio bands corresponds to the peak in VHE emission. We explain this anti-correlation in terms of a possible leptonic origin of the observed VHE $\gamma$-ray emission. We determine the spectral indices of the bright individual filaments, 
which  were detected in both the 610 and the 325 MHz bands.  Our values range from $-$1.11 to $-$0.15, consistent with the non-thermal radio emission. We also report a possible radio counterpart of a nearby TeV source HESS J1729$-$345 from the 843 MHz Molonglo Galactic Plane Survey and the 1.4 GHz Southern Galactic Plane Survey maps. The positive radio spectral index of this possible counterpart suggests a thermal origin of the radio emission of this nearby TeV source.
\end{abstract}

\begin{keywords}
ISM: supernova remnants -- ISM: individual objects: SNR G353.6$-$0.7  -- radio continuum: ISM -- radiation mechanisms: non-thermal -- acceleration of particles -- cosmic rays 
\end{keywords}



\section{Introduction}

Galactic cosmic rays are energetic particles believed to be mostly composed of hadrons with energies ranging from $\sim$ 0.1 to 10$^{8-9}$ GeV \citep{blandford1987}. How are these particles accelerated to these high energies remains a fundamental question. Supernova remnant (SNR) shocks are considered as one of the 
probable candidates of these particle acceleration sites \citep{blasi2013}.  
Cherenkov telescopes  such as the High Energy Stereoscopic System (H. E. S. S.), the Major Atmospheric Gamma Imaging Cherenkov Telescope (MAGIC) and the Very Energetic Radiation Imaging Telescope Array System (VERITAS) have discovered many very high energy (VHE; Energy $>$ 100 GeV)
 $\gamma$-ray sources from the Galactic plane in the  last decade \citep{degrange2015}.  Association of  spatially-resolved shell-like  $\gamma$-ray 
 sources with SNRs is the signature of particle acceleration at supernova shocks. 
 Studying these objects provides a unique opportunity to probe the cosmic-ray particles that emit high energy $\gamma$-rays. However, so far there are only five spatially-resolved shell-like VHE sources firmly identified as SNRs \citep[for a review, see ][]{acero2015}. These are RX J1713.7$-$3946 \citep{Aharonian2004,hess2016a}, RX J0852.0$-$4622 \citep{Aharonian2007,hess2016b}, RCW 86 \citep{Aharonian2009,hess2016c}, SN 1006 \citep{Acero2010} and HESS J1731$-$347 \citep{Abramowski2011}.

HESS J1731$-$347 was discovered \citep{Aharonian2008} as a VHE $\gamma$-ray source in the H. E. S. S. Galactic Plane Survey at $\alpha_{\rm J2000}$ = 17$^{\rm h}$31$^{\rm m}$55$^{\rm s}$, $\delta_{\rm J2000}$ = $-$34$^{\circ}$42$^{\prime}$36$^{\prime \prime}$ with no identified counterpart in other wavebands. Later \citet{Tian2008} discovered a faint shell-type radio source SNR G353.6$-$0.7 with an angular size of $\approx$ 30 arcmin,  in spatial coincidence with HESS J1731$-$347 
in the Southern Galactic Plane Survey (SGPS) at 
1420 MHz \citep{Haverkorn2006} and Molonglo Galactic Plane Survey (MOST) at 843 MHz \citep{green1999}. \cite{Tian2008} derived an integrated flux density of 2.2$\pm$0.9 Jy at 1420 MHz by azimuthally averaging the radio intensity in rings about the centre of the SNR. 
Later a deeper $\gamma$-ray observation of HESS J1731$-$347 revealed the shell-like morphology in $\gamma$-rays as well \citep{Abramowski2011}, which made HESS J1731$-$347 the fifth member of the VHE shell SNR group. HESS J1731$-$347 was also observed with the {\it Fermi} Large Area Telescope (LAT), but no  GeV counterpart was detected \citep{yang2014}. 
 HESS J1731$-$347 was detected in X-rays in archival {\it ROSAT} data  \citep{Aharonian2008,Tian2008}. The X-ray emission was found to be non-thermal  \citep{Tian2010,Abramowski2011,bamba2012}, indicating that   the electron population producing these 
non-thermal X-rays are of TeV energies. A compact object XMMU J173203.3$-$344518 was detected at the centre of the remnant  by \cite{halpern2010a} 
using {\it XMM-Newton} archival data. This object was considered as a central compact object (CCO) associated with SNR G353.6$-$0.7 \citep{acero2009, halpern2010a, Tian2010}. 
\cite{Tian2008} estimated the distance to the SNR as 3.2$\pm$0.8 kpc, by assuming the nearby \ion{H}{ii} region G353.42$-$0.37 to be at the same distance as the SNR.

In this paper, we present the Giant Metrewave Radio Telescope \citep[GMRT, ][]{swarup1991} observations of SNR G353.6$-$0.7 in the 325, 610 and 1390 MHz bands.
Our detections in the 325 and 610 MHz bands are the lowest frequency detections of the SNR thus far. 
 We compare the shell morphology of the SNR from our low-frequency measurements with those of other published high-frequency results. We also measure the spectral index of different filaments of the SNR.
The paper is organised as follows: In \S 2 we explain the GMRT observations  and in \S 3 we briefly discuss the method of data analysis. 
The \S 4 contains the results and discussions and the \S 5 contains the summary and conclusions.

\section{GMRT Observations}

We observed SNR G353.6$-$0.7 with the GMRT at 325, 610 and 1390 MHz. The 325, 610 and 1390 MHz observations were carried out on 2013 Oct19.3 (UT), 2015 Aug 29.5 (UT) and 2013 Dec 19.1,20.1 (UT), respectively. In the 325 and 610 MHz bands, the GMRT full width at half maximum (FWHM) of primary beam is 85.2 and 43 arcmin, respectively, and only one pointing was necessary to cover the SNR. In
the 1390 MHz band, the FWHM is 26.2 arcmin, and observations were taken in four pointings since the field-of-view (FOV) is less than the source extent. Three pontings were done towards  the SNR region and one pointing  towards another TeV source HESS J1729$-$345, just 30 arcmin away from the SNR \citep{Aharonian2008}. The 1390 MHz observations were carried out on two consecutive days so as to increase the sensitivity. Details of the GMRT observations are given in Table \ref{tab:observation}.

The data was recorded with an  integration time of 16.1 s. For all the frequency bands, the bandwidth chosen was 33 MHz split into 256 frequency channels. All the observations were taken with Automatic Level Control (ALC) on. The ALC controls the amplitude gains of the antennas with a negative feedback loop at the output of each antenna. It ensures a safe operating point without running the electronics into saturation. 
3C286 was used as the flux density calibrator at all frequencies. At 325 MHz, sources J1714$-$252 and J1830$-$360 were used as the phase calibrators.  
J1830$-$360  and J1714$-$252 were used as phase calibrators at the 610 and 1390 MHz observations, respectively. The flux density calibrator was used to calibrate antenna gains, and the phase calibrators were used to correct for phase variations due to atmospheric fluctuations. The phase calibrators were also used for bandpass calibration.

 The system temperature ($T_{\rm sys}$) correction was implemented by determining the self-powers of each antenna with ALC off in a separate test observation run,
taken in December 2015 in the 325, 610 and 1390 MHz bands. The SNR and the flux calibrator were observed for 10 minutes each.  Power equalisation was done on the SNR. Antenna self-powers (total power measured by antenna) were recorded for all antennas, for each time stamp (16.1 s) and for all channels. 

\begin{table*} 
	\caption{Details of GMRT observations.}
	\label{tab:observation}
	\begin{tabular}{lccccc} 
    \hline
	
	Frequency & Date of observation & Time on source & FWHM of primary beam&  Number of & Pointing centre \\
	 (MHz)     & (UT) & (hours) & (arcmin) & pointings & RA (J2000) Dec(J2000)\\
	\hline
 325 & 2013 Oct 19.3  & 5.4 & 85.2 & 1 & 17$^{\rm h}$ 31$^{\rm m}$ 54.9$^{\rm s}$   $-$34$^{\circ}$ 42$^{\prime}$ 36.00$^{\prime \prime}$ \\
 \hline
	
    610 & 2015 Aug 29.5  &  3.5 & 43 & 1  &  17$^{\rm h}$ 32$^{\rm m}$ 19.6$^{\rm s}$  $-$34$^{\circ}$ 43$^{\prime}$ 30.00$^{\prime \prime}$\\
    \hline
    
	1390$^{*}$ & 2013 Dec 19.1, 20.1  & 2.5  & 26.2 &  4 & 17$^{\rm h}$ 32$^{\rm m}$ 11.9$^{\rm s}$  $-$34$^{\circ}$ 37$^{\prime}$ 60.00$^{\prime \prime}$\\
	
	 &   &  & &  & 17$^{\rm h}$ 32$^{\rm m}$ 29.9$^{\rm s}$  $-$34$^{\circ}$ 49$^{\prime}$ 60.00$^{\prime \prime}$\\
	
	&  &  &  &   & 17$^{\rm h}$ 31$^{\rm m}$ 35.9$^{\rm s}$  $-$34$^{\circ}$ 44$^{\prime}$ 60.00$^{\prime \prime}$\\
	
	 &  & &  &   & 17$^{\rm h}$ 29$^{\rm m}$ 35.9$^{\rm s}$  $-$34$^{\circ}$ 32$^{\prime}$ 60.00$^{\prime \prime}$\\
		\hline
\multicolumn{6}{l}{*In the 1390 MHz band, observations were taken in two consecutive days with 4 pointings to cover SNR G353.6$-$0.7 and HESS J1729$-$345.}	
\end{tabular}
\end{table*}

\section{Data Analysis}

\subsection{Calibration and Imaging}

The data was analysed using Astronomical Image Processing System (AIPS) developed by the National Radio Astronomy Observatory \citep[NRAO, ][]{Greisen2003}. One antenna was not working in the 325 as well as 610 MHz band observations whereas three
 antennas were not working in the 1390 MHz observations due to technical problems. These antennas were excluded from further analysis. Corrupted data due to instrumental problems were removed using standard AIPS routines. Channels with radio frequency interference (RFI) were flagged using the task SPFLG \footnote{Detailed documentation of SPFLG as well as all other AIPS  tasks mentioned in this section is available at http://www.aips.nrao.edu/cook.html}. This task displays the $uv$ data in a grid with frequency channels in the x axis and time in the y axis. It also provides interactive options to inspect and flag the data. Amplitude and phase calibration were done for single channel and the solutions were applied to all the channels in the data. Bandpass calibration was done using the phase calibrator. At 1390 MHz band, the two days  of data were combined together after calibration using the AIPS task DBCON. The target source data was averaged in small frequency bins so as to reduce the effect of bandwidth smearing, which reduces the amplitude of the visibility. The fractional reduction in the strength of a source at a radial distance $r$ from the centre of field to that at the centre of field is given by
\begin{equation}\nonumber
R_{\rm b}= 1.064 \frac{\theta_{\rm b}\nu_{0}}{r\Delta\nu}{\rm erf}\left(0.833 \frac{r\Delta\nu}{\theta_{\rm b}\nu_{0}}\right) 
\end{equation}
where $\theta_{\rm b}$ is the angular size of synthesized beam, $\nu_{0}$ is the centre of the observing band and $\Delta \nu$ is the bandwidth of the signal. We averaged a few channels in such a way that the fractional reduction in the strength of a source at the edge of the field is 5 per cent. For example, in the 325 MHz band, the adjacent 4 channels were averaged to obtain a channel width of 0.5 MHz and 8 channels were averaged to get a channel width of 1 MHz in the 610 MHz band. In the 1390 MHz, 10 channels were averaged to obtain a channel width of 1.25 MHz. These sub-channels were stacked together while making images. 
 
Multi-facet imaging was done using the task IMAGR as the  two dimensional fast fourier transform leads to a flat sky approximation. 
Multi-facet imaging takes care of the phase errors due to this approximation. The number of facets was calculated using the AIPS task SETFC. A total of 55 facets were created for the 325 MHz data, and 37 facets in the 610 MHz band. In the 1390 MHz band, 31 facets were created  for each pointing.
First, high-resolution images of the compact sources were made for self-calibration using visibilities from baselines greater than 1 km, i.e. excluding the
GMRT central square antennas. 
The images were made using \textit{uv} range 2 to 25~k$\lambda$, 3 to 30 k$\lambda$  and 5 to 105 k$\lambda$ in the 325, 610 and 1390 MHz bands, respectively. Only point sources were cleaned, whereas, the SNR region and other extended sources were not cleaned during high-resolution imaging.  
A few rounds of phase only self-calibration and two rounds of amplitude and 
phase self-calibration were run. All the clean components were subtracted from the \textit{uv} data using task UVSUB, which takes the $uv$ data set and clean component files as inputs and subtracts the clean components from the $uv$ data set. This procedure was followed essentially to remove the compact sources in the field since  the SNR has extended structure.

Finally,  low-resolution maps were made using small \textit{uv} range (only central square data).
This corresponded to a \textit{uv} range of 0 to 2 k$\lambda$, 0 to 3 k$\lambda$ and 0 to 5 k$\lambda$ for the 325, 610 and 1390 MHz bands data, respectively. The shortest baselines we have for these observations are $B_{\rm min}$ $\sim$ 50, 100 and 500$\lambda$ for the 325, 610 and 1390 MHz bands respectively. In the 1390 MHz band, a significant fraction of short baselines was flagged due to RFI. This makes the interferometer sensitive to angular scales less than 0.6$\lambda/B_{\rm min}$ \citep{Basu2012}, provided there are enough measurements at short baselines. All facets in each pointing were combined using task FLATN. This task does an interpolation of multiple fields produced by the task IMAGR into a single image. In the 1390 MHz band, the final map was made by combining maps of all three pointings towards the SNR. Finally primary beam corrections were made for all the maps. 

Our observations are sensitive up to the largest angular scales of $\sim$ 41.2, 20.6 and 4.0 arcmin in the 325, 610 and 1390 MHz bands, respectively. Since the size of the SNR G353.6$-$0.7 is 30 arcmin, we do not have enough short spacings to completely measure the flux density of SNR at the 610 and 1390 MHz frequencies. However, the effect of missing flux is small in the 325 MHz map. 
      A simulation was done to quantify the flux density missing using AIPS task UVMOD. This task allows to modify the $uv$ data by adding a model. A sky model of a solid disc of 30 arcmin in diameter was simulated and then sampled to the GMRT $uv$-coverage at the 325 MHz. The $uv$-data was then imaged and more than 97 per cent of flux density was recovered. Thus our GMRT 325 MHz map suffer from a maximum of 3 per cent  missing flux while imaging a 30 arcmin structure. Simulations for missing flux were done for individual filaments at 610 MHz which we describe in \S 4.3.
      
The SNR was detected in the 325 and 610 MHz bands (see Fig. \ref{maps}); however, we do not detect it in the 1390 MHz band because of the missing flux due to lack of short spacings and the sensitivity limitation.

\subsection{System Temperature ($T_{\rm sys}$) correction}

The SNR G353.6$-$0.7 is close to the Galactic plane, where the sky temperature ($T_{\rm sky}$) is high due to the contribution of the Galactic diffuse emission. Since observations were done with the ALC ON in its  default mode and ALC controls the antenna gains automatically, the antenna gains are reduced with respect to the calibrators and the flux density measurements will be less than their true value. As the antenna self-powers are proportional to the system temperatures ($T_{\rm sys}$), a correction can be found by measuring the ratio of antenna self-powers towards the target source and flux calibrator, 
under the assumption that the contribution of electronics towards the system temperature will not change significantly between the days of observations at the same frequency. 
The actual flux density of the target source can be recovered by applying this $T_{\rm sys}$ correction. 

Test observations for the $T_{\rm sys}$ correction were taken in December 2015, and the data was analysed using GMRT specific offline packages. 
After removing corrupted data, the ratio of self-powers was found for each antenna. The mean and the rms of the ratios were computed to estimate the 
average  $T_{\rm sys}$ correction. A detailed explanation of $T_{\rm sys}$ correction can be found in GMRT technical report by 
Roy (2006)\footnote{http://ncralib1.ncra.tifr.res.in:8080/jspui/handle/2301/315}.

 Another way of obtaining $T_{\rm sys}$ correction is to estimate $T_{\rm sky}$ at both target source and flux calibrator positions from the all-sky temperature map. Haslam 408 MHz full sky temperature map by \citet{Haslam1982} and 150 MHz temperature measurements by \cite{Landecker1970} were used to obtain $T_{\rm sky}$ at both positions. The $T_{\rm sky}$ at 325, 610 and 1390 MHz were calculated by assuming a spectral index of $-$2.55 for Galactic diffuse emission \citep{roger1999}. System temperature has contributions from the sky, ground and receiver. The ground temperature ($T_{\rm g}$) and receiver temperature ($T_{\rm rec}$) were taken from a GMRT technical report\footnote{http://gmrt.ncra.tifr.res.in/gmrt$_{-}$hpage/Users/doc/manual\\/Manual$_{-}$2013/manual$_{-}$20Sep2013.pdf}.   

The $T_{\rm sys}$ correction obtained for each frequency from GMRT test observations, from the Haslam map \citep{Haslam1982} and from the 150 MHz map \citep{Landecker1970} are given in Table \ref{tsys}. The $T_{\rm sys}$ correction from the Haslam map is 54.6 per cent greater than that obtained from GMRT test observations at 325 MHz and 27.6 per cent greater at 610 MHz. The correction from the 150 MHz map is 39.7 per cent greater than that of the GMRT correction factor at 325 MHz and 17.0 per cent greater at 610 MHz. At 1390 MHz, all three correction factors are similar. A similar trend in $T_{\rm sys}$ corrections with GMRT observations and the Haslam map was also reported by \cite{marcote2015}.

The $T_{\rm sky}$ measurements using GMRT at 240 MHz was determined by \cite{sirothia2009}. They report differences in the $T_{\rm sky}$ measurements obtained from the Haslam map and GMRT measurements, and the rms of percentage differences is 56 per cent. This discrepancy might be due to the assumption of a constant spectral index of $-$2.55 across the whole sky, while determining $T_{\rm sky}$ using the Haslam map. This assumption need not hold in all regions of the sky, especially towards the Galactic centre where significant contribution from thermal emission is expected. Moreover, variation in $T_{\rm rec}$ with elevation and ambient temperatures  \citep{sirothia2009}, will also affect the $T_{\rm sys}$ measurements. Estimation from the Haslam map does not account for these issues. Hence $T_{\rm sys}$ correction from GMRT test observation seems to be more reliable than the one obtained from the Haslam map. In this paper, we use the correction factor from GMRT observations to scale the observed flux density values.

\begin{table*} 
	\caption{$T_{\rm sys}$ correction from different methods.}
	\label{tsys}
	\begin{tabular}{lccccr} 
    \hline
	
	Frequency & Correction & Correction using 408 MHz map & Correction using 150 MHz map\\
	   (MHz)  & (GMRT) & (Haslam et al. 1982) & (Landecker \& Wielebinski, 1970) \\
	\hline
	325 & 4.01$\pm$0.51  & 6.20 & 5.60 \\
    610 & 1.59$\pm$0.13 & 2.03 & 1.86 \\
	1390 &1.08$\pm$0.04 & 1.10 & 1.10\\
		\hline
		 
	\end{tabular}
	
	\scriptsize{Notes: Corection (GMRT) is the mean of ratios of antenna self-powers measured towards SNR and flux calibrator. The uncertainities are at 1$\sigma$ level. Correction (Haslam et al. 1982) and correction (Landecker \&  Wielebinski, 1970) are corrections obtained from sky temperature maps.}
\end{table*}

\subsection{Spectral Analysis}

\begin{figure*}

	\includegraphics[width=\columnwidth]{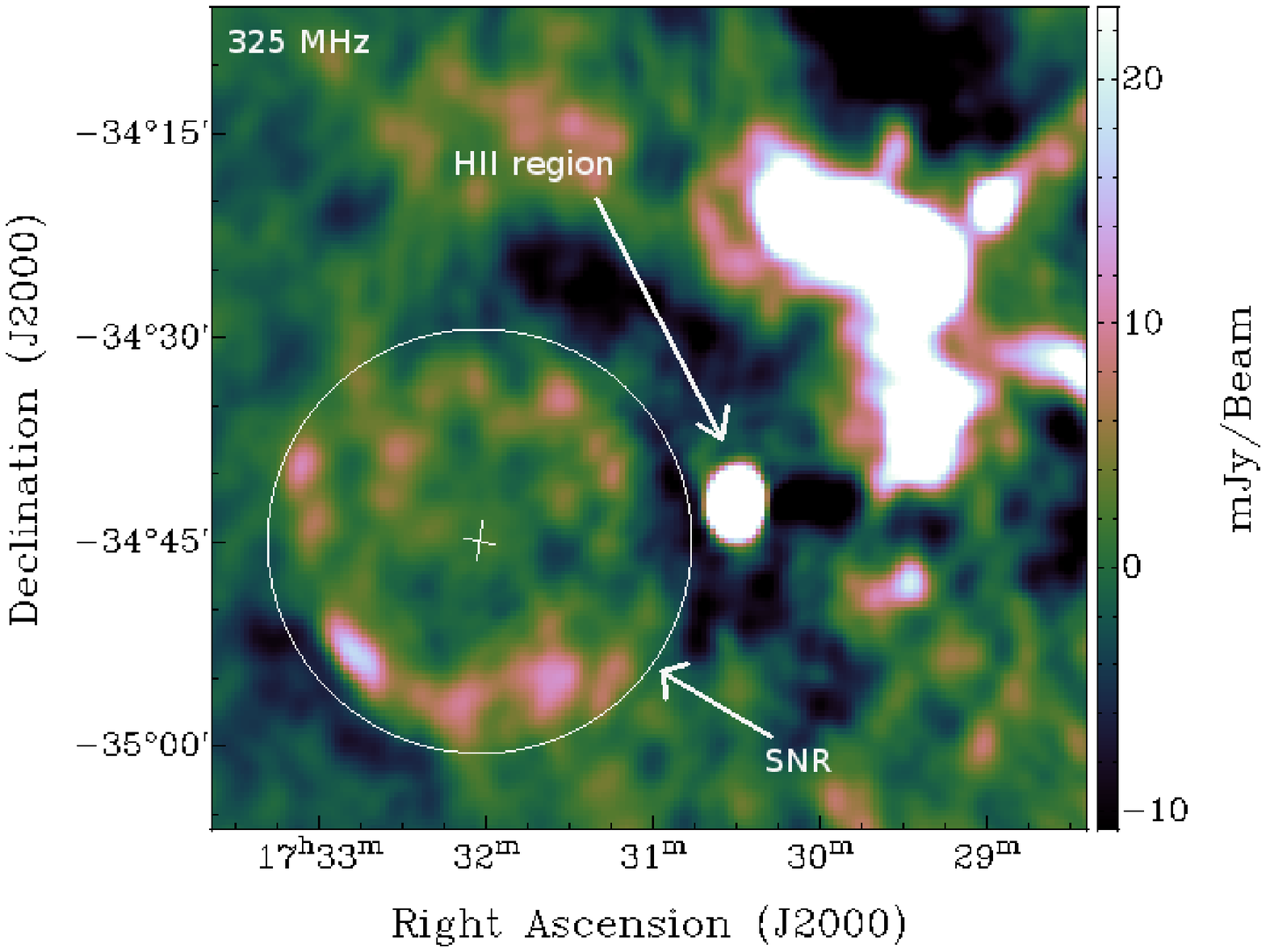}
	\includegraphics[width=\columnwidth]{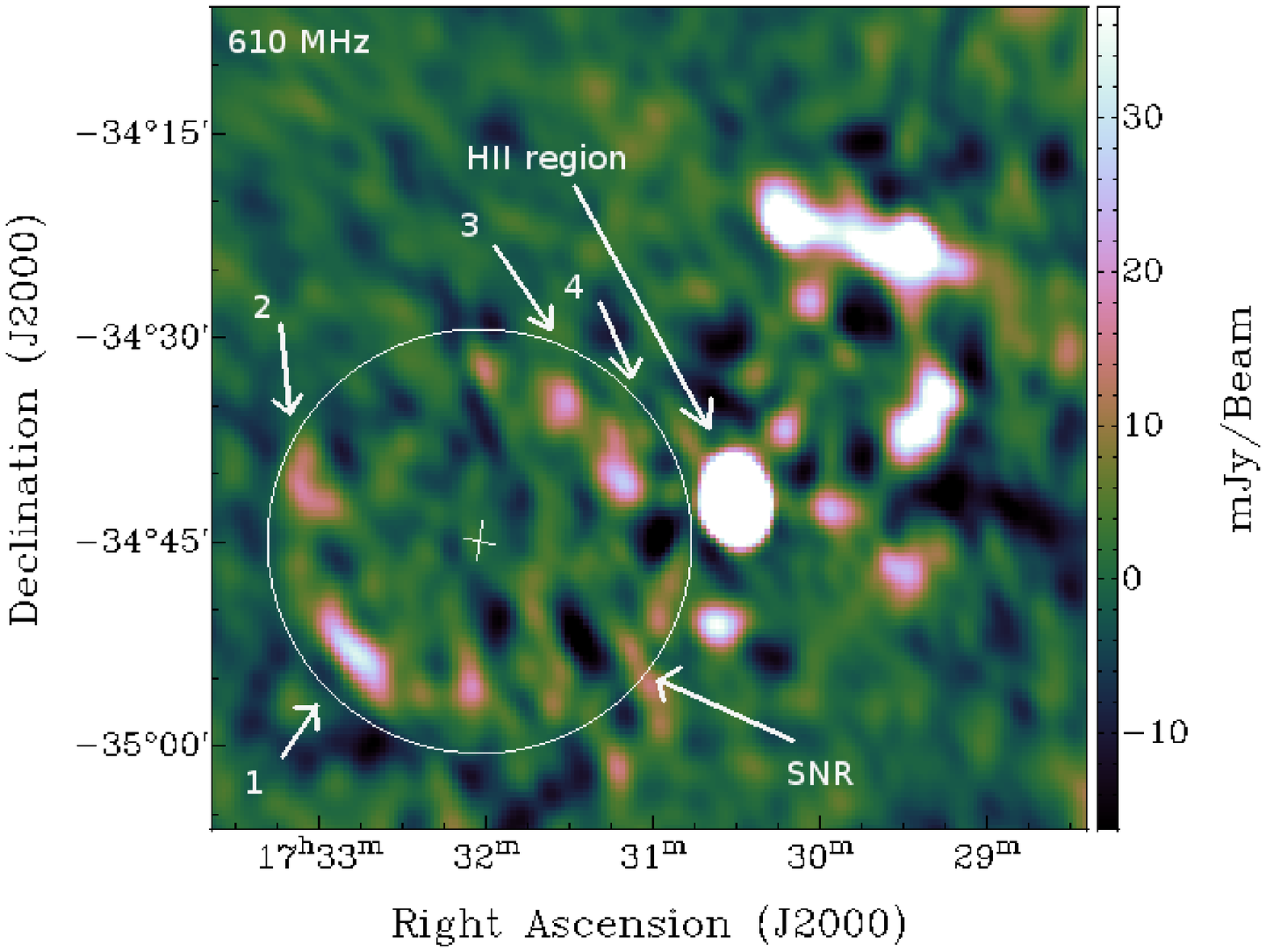}
	\includegraphics[width=\columnwidth]{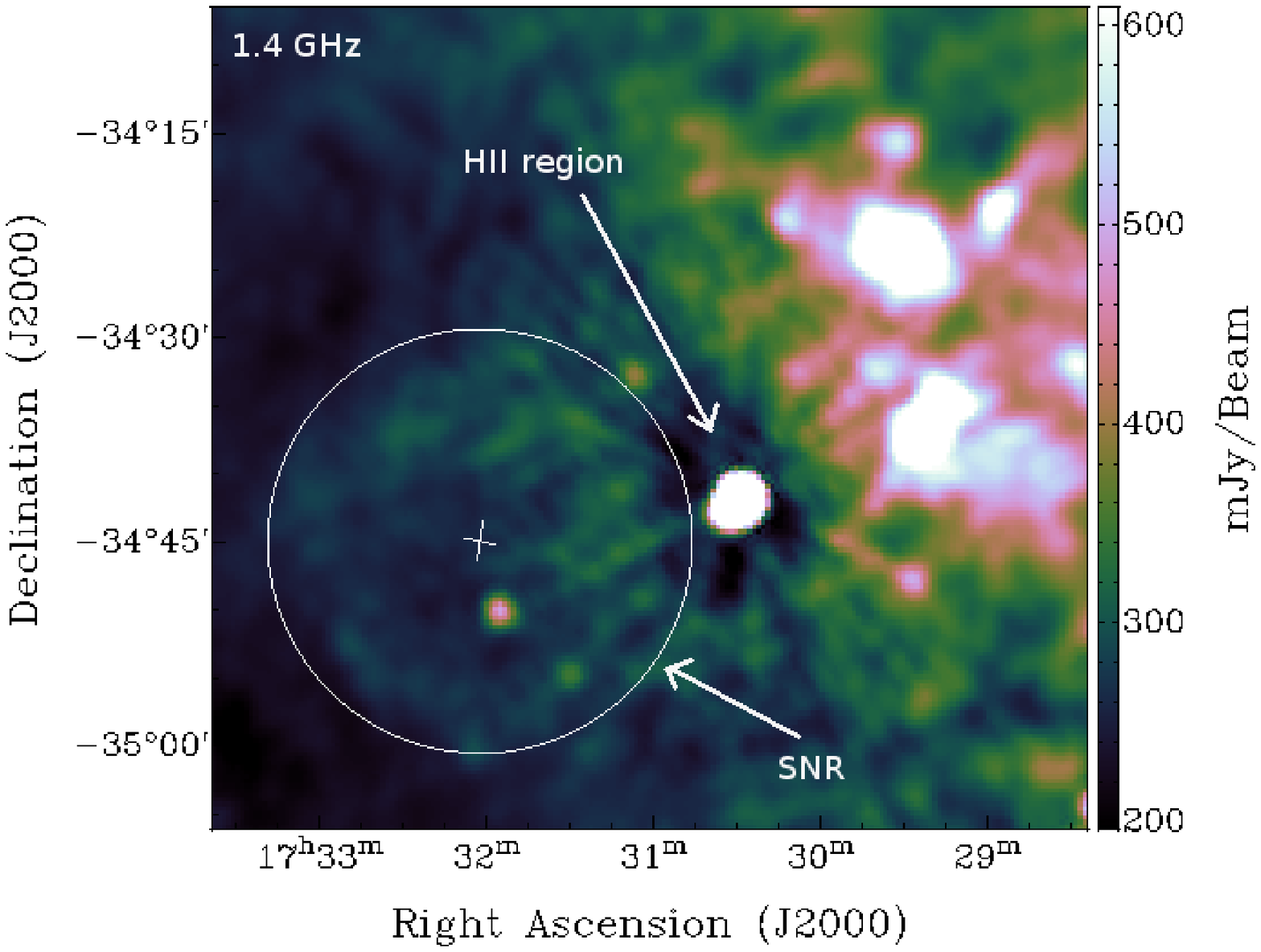}
	\includegraphics[width=\columnwidth]{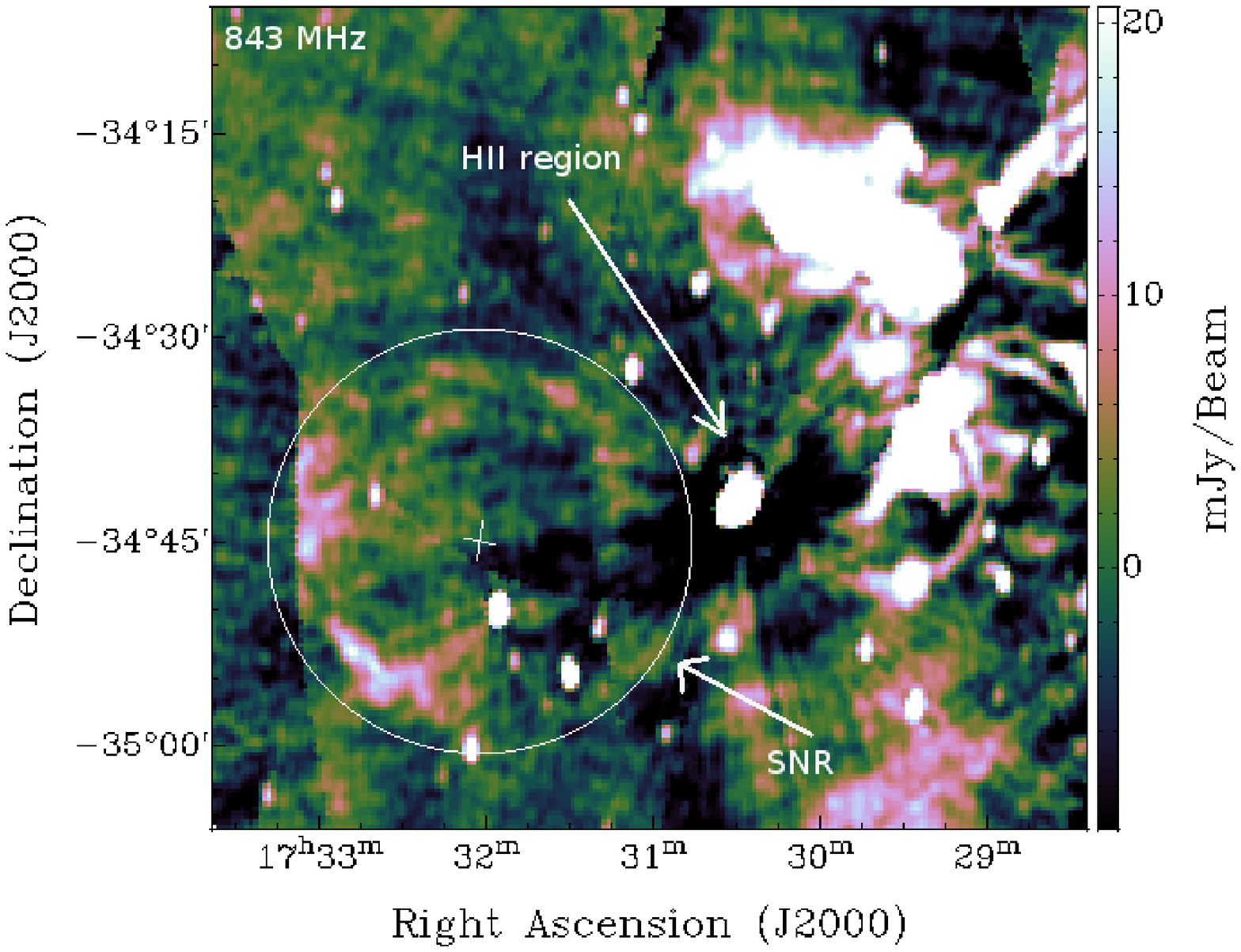}
    \caption{{\it Top left}: The GMRT low resolution map of the field containing SNR G353.6$-$0.7 at 325 MHz. Map resolution is 135$\times$97 arcsec$^{2}$, rms=4.8 mJy beam$^{-1}$. {\it Top right}: we show GMRT low resolution map of the field containing SNR G353.6$-$0.7 at 610 MHz. Map resolution is 150$\times$105 arcsec$^{2}$, rms=4 mJy beam$^{-1}$. Here the areas marked by 1, 2, 3, 4 are the filaments for which spectral indices were estimated. Compact sources have been removed from both 325 and 610 MHz GMRT maps. For comparison, we show the SGPS map \citep{Haverkorn2006} of the field containing SNR G353.6$-$0.7 at 1420 MHz with resolution is 100$\times$100 arcsec$^{2}$, rms=10 mJy beam$^{-1}$ ({\it bottom left }),
    as well as the MOST map \citep{green1999} of the field containing SNR G353.6$-$0.7 at 843 MHz. with resolution is 49.71$\times$43.00 arcsec$^{2}$, rms=3 mJy beam$^{-1}$ 
    ({\it bottom right}). The circles in all four images are of same size and centred at the CCO position (marked with a cross) indicate the area enclosing SNR G353.6$-$0.7. The colour scheme used in the maps is cubehelix \citep{Dave2011}.
    }
    \label{maps}
\end{figure*}

Analysis of spectral index ($\alpha$, where the radio flux density $S_{\nu}$ at frequency $\nu$ is related to $\alpha$ as $S_{\nu}$= $\nu^{\alpha}$)
provides information about the nature of the radio emission. Since the full shell structure was not detected in the 610 MHz band, we determined spectral indices  for four filaments (see filaments marked 1, 2, 3 and 4 in the top right of Fig. \ref{maps} and Table \ref{filaments})  which were detected in both the 610 and 325 MHz bands. The sizes of the filaments are less than 11 arcmin and are well below the maximum angular scales that can be mapped with GMRT in both the 325 and 610 MHz bands. We estimated a missing flux of less than 2 per cent for these filaments in both the 325 and 610 MHz bands in our observations. Thus these structures are not affected from significant missing flux problems.

To determine the spectral indices, both 325 and 610 MHz maps were convolved to the coarser resolution (150$\times$105 arcsec$^{2}$) using the AIPS task CONVL. Both maps were then aligned to same geometry using the AIPS task OHGEO. 
The flux density of the same region over both maps was determined using task BLSUM and rescaled with the $T_{\rm sys}$ correction factor to get the actual flux density. The total error $\sigma_{\rm total}$ in the flux density measurement is a combination of errors from calibration and rms noise of the map. Hence the uncertainties in flux density measurements are estimated as
\begin{equation}
\sigma_{\rm total} = \sqrt{\sigma_{\rm rms}^{2} + \sigma_{\rm cal}^{2}}
\end{equation}
where $\sigma_{\rm rms}$ is the map rms noise and $\sigma_{\rm cal}$ is the calibration error. We assumed that the calibration errors are 10 per cent of the measured flux density. The errors on spectral indices are computed using standard error propagation formulas.

\section{Results and Discussion}

\subsection{Morphology of SNR G353.5$-$0.7}

Our observations with the GMRT have revealed the radio morphology of the SNR  at the 325 and 610 MHz frequencies. Fig. \ref{maps}  (top panels) clearly shows a shell-like structure with filaments. The extent of shell structure is $\thicksim$ 30 arcmin, consistent with that measured by \cite{Tian2008} from the SGPS map at 1.4 GHz \citep[bottom left panel of Fig. \ref{maps}, ][]{Haverkorn2006}. The 325 MHz map has an rms of 4.8 mJy beam$^{-1}$ with a resolution of 135$\times$97 arcsec$^{2}$. Since the SNR is extended, low-resolution maps were made in order to get better signal to noise ratio per synthesized beam. The radio emission is brightest in the filament seen towards the south-east, and is detected at 10$\sigma$ significance. The faintest filament is towards the north and is detected at 3$\sigma$ significance.
The rms of the 610 MHz map is 4 mJy beam$^{-1}$  for a resolution of 152$\times$105 arcsec$^{2}$. The diffuse faint emission towards the north is not detected in the 610 MHz band. This is due to sensitivity limitation and poor sampling at short baselines. The smaller field of view at the 610 MHz (43 arcmin) also limits better mapping of the 30 arcmin structure in the sky. Filaments 1 and 2 (top right panel of Fig. \ref{maps}) seen in the 610 MHz GMRT map are also seen in the MOST 843 MHz map \citep[bottom right panel of Fig. \ref{maps}, ][]{green1999}. The brightest emission comes from the south-east filament (Filament 1), which is also the case in the GMRT 325 MHz image.

The SNR is not detected in the 1390 MHz GMRT map. The map rms is 1.8 mJy beam$^{-1}$. The shortest baseline available with the GMRT at 1390 MHz is $\sim$ 500$\lambda$. This makes the interferometer sensitive to a maximum angular scale of 4 arcmin. Thus it is not possible to image the full SNR shell at 1390 MHz using the GMRT. But the bright filamentary structures of small angular size can be detected. However, we did not detect any filaments due to sensitivity limitation and lack of enough short baselines. The flux densities of point sources in the GMRT map are compared with the flux densities from NRAO VLA Sky Survey (NVSS) map at 1.4 GHz and they matched within 10 -- 15 per cent. The brightest filament seen in both the 325 and the 610 MHz GMRT maps is also seen  as a 4$\sigma$ feature in the NVSS map which has an rms of 0.7 mJy beam$^{-1}$.

\citet{Tian2010} overlaid the  X-ray map with the 1.4 GHz SGPS contour map, which  revealed that the X-ray  emission extends beyond the radio shell  towards the northern region \citep[see Fig. 1 of][]{Tian2010}. This is not expected since X-rays are found to be non-thermal  \citep{Tian2010,Abramowski2011,bamba2012}. The non-thermal X-ray and radio emission is expected to come from the synchrotron radiation from relativistic electrons accelerated at the shock. However, our 325 MHz GMRT maps clearly show radio emission to be present in this region also, eliminating this discrepancy.

The bright compact source G353.464$-$0.690 \citep{Zoonematkermani1990} inside the remnant which is seen in the 1.4 GHz SGPS image is not seen in the GMRT map. This is because we have  subtracted point sources during our analysis. The bright radio source J173028$-$344144 \citep{Condon1998} seen towards the west of the SNR is a \ion{H}{ii} region. The extended structure seen further west is also likely to be a \ion{H}{ii} region since both of these are seen in the IRAS 60$\upmu$m map \citep{Neugebauer1984} and the WISE 22$\upmu$m map \citep{Wright2010}. 

The integrated flux density of the SNR at the 325 MHz is 1.84 $\pm$ 0.15 Jy. For a typical SNR spectral index of $\alpha$ $\sim$ $-$0.5 \citep{jones1998}, the flux density of the SNR at 1420 MHz is expected to be 0.88 Jy. \cite{Tian2008} derived 1420 MHz flux density using combined ATCA and Parkes telescope data to be 2.2$\pm$0.9 Jy. This value is likely to be an overestimation due to the contamination from thermal emission from nearby \ion{H}{ii} region. This discrepancy may also be due to the flux density estimation made by \cite{Tian2008} by taking one-half of the SNR (half towards low Galactic latitude) and extrapolating to the other half to estimate the complete flux density. However, our values match within 1.5$\sigma$ with that of \cite{Tian2008}. The change in flux density in the low-frequency end will change the shape of the spectral energy distribution (SED) and the parameters derived from multi-wavelength models.

\subsection{Central compact object}
A bright compact object , XMMS J173203$-$344518, was discovered in the X-ray band near the centre of the SNR by \cite{Tian2010} from the {\it XMM} observations. The CCO was also seen in the {\it Suzaku} and the {\it Chandra} data \citep{halpern2010a}. A marginal pulsation period of 1 s was found for the XMMS J173203$-$344518, which is between the periods of magnetars ($2 -12$ s) and the CCOs ($0.1 - 0.4$ s) \citep{halpern2010a}. However, \cite{halpern2010b} failed to confirm this pulsation with the later {\it Chandra} observations. There is no evidence for a radio counterpart for this object \citep{Tian2010}. We did not detect radio emission at the CCO position in the GMRT observations at any frequencies. We provide deep limits on any possible radio counterpart of the CCO. The 3$\sigma$ flux density upper limits at the CCO position are 727, 780 and 184 $\mu$Jy at the 325, 610 and 1390 MHz, respectively, from our GMRT high-resolution maps.

\subsection{Spectral Index}
\label{sec:spec}

The global radio spectral index of the SNR cannot be determined since the complete shell is not detected at the 610 MHz. We determined the spectral indices of four filaments of the SNR labelled 1 to 4 in the Fig. \ref{maps}. The physical dimensions in angular units and flux densities at the 325 and 610 MHz of the filaments are listed in Table \ref{filaments}. The sizes of the filaments are less than 11 arcmin with  less than 2 per cent missing flux at both the 610 and the 325 MHz  observations. We have also accounted for the background contribution to the SNR flux density. However, these are statistical estimates, and the contribution of background flux density at the SNR position is uncertain. 

The spectral index values are $-$0.70 $\pm$ 0.19, $-$1.11 $\pm$ 0.22, $-$0.5 $\pm$ 0.3 and $-$0.15 $\pm$ 0.32 for the filament 1, 2, 3 and 4 respectively. The spectral index values are broadly consistent with the non-thermal emission. The spectral index of the \ion{H}{ii} region is found to be 0.63$\pm$0.14 which is consistent with thermal emission.

Here we note a steeper spectral index $-1.11\pm0.22$ for the filament 2.  This steepening is consistent with synchrotron cooling which steepens the spectrum by a factor of $\Delta \alpha=0.5$ from normal synchrotron spectral index. If this steepening is real, this could have important implications for the magnetic field in the SNR shell, and thus the VHE emission scenario (see \S \ref{comparison}).  Our future observations with the upgraded GMRT with 200 MHz bandwidth  and 2.5 times higher sensitivity compared to the existing GMRT will be very crucial here to determine the precise spectral indices with much improved error bars. The wide bandwidth of upgraded GMRT provides excellent $uv$ coverage, thus more sensitivity to faint extended radio emission, which is crucial in accurately determining 
the spectral indices of the filaments. 
 
\begin{table*} 
	\caption{Physical dimensions of four filaments in angular units and flux densities at 325 MHz and 610 MHz.}
	\label{filaments}
	\begin{tabular}{cccccc} 
    \hline
	
	Filament & Length  & width & S$_{\rm int}$ (325 MHz) & S$_{\rm int}$ (610 MHz)& Spectral index  \\
	     & (arcmin) & (arcmin) & (mJy) & (mJy)  & \\
	\hline
	Filament 1 & 11 & 3 & 106.22 & 68.67   & $-$0.70$\pm$0.19 \\
    Filament 2 & 6 &  2.8 & 121.66 & 58.94 & $-$1.11$\pm$0.22\\
	Filament 3 & 5.4 & 3.7 & 56.90 & 41.55 & $-$0.50$\pm$0.30\\
	Filament 4 & 5.9 & 3.4 & 42.99 & 39.45 & $-$0.15$\pm$0.32\\
		\hline
	\end{tabular}
	
	\scriptsize{S$_{\rm int}$ denotes the integrated flux density of the filaments at each frequencies.}
\end{table*}

\subsection{Comparison with the VHE emission}
\label{comparison}

\begin{figure}
	\includegraphics[width=9 cm]{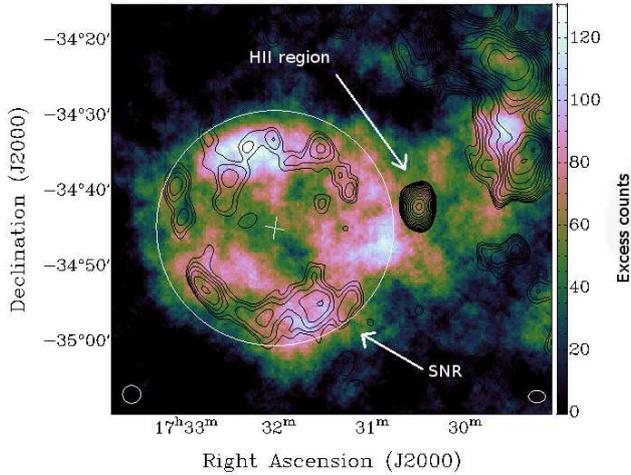}
    \caption{TeV $\gamma$-ray excess map \citep{Abramowski2011}  (resolution 144$\times$144 arcsec$^{2}$; shown in the bottom left corner of the image) overlaid with GMRT 325 MHz map (resolution 135$\times$97 arcsec$^{2}$; shown in the bottom right corner of the image) contours. The scale of the $\gamma$-ray excess map is excess counts per smoothing gaussian width ($\sigma$ = 0.04 deg). The contours are $\pm\sqrt{2}^{n}\sigma$ mJy beam$^{-1}$ ($n = -2,2,3,4,5,6;$ $\sigma$=1.5 mJy beam$^{-1}$). The circle indicates the area enclosing SNR G353.6$-$0.7 centred at the CCO position (marked with a cross). The colour scheme used in the map is cubehelix \citep{Dave2011}}
    \label{map2}
\end{figure}

Since the structure of the SNR is clearly detected in the 325 MHz radio map, the spatial correlation of the radio emission with that of the VHE emission  can be compared.
We compared the sites of synchrotron radio emission with the VHE emission measured with the H. E. S. S. \citep{Abramowski2011}, as shown in Fig. \ref{map2}. The spatial extent of the radio emission is similar to that of the VHE emission except for the region towards the west, where there is no radio emission though bright VHE emission is seen. 
It is likely that the radio emission in this region is  sensitivity limited as this region has higher rms, due to contribution from contamination from nearby sources. The 3$\sigma$ upper limit of flux density in this region at 325 MHz is 63 mJy which is comparable to the flux densities of filaments 3 and 4 (see Table \ref{filaments}). The morphology of the SNR is shell-like in both radio and VHE emission, but the finer structure do not seem to be correlated. The bright filaments in radio (south-east filament and eastern filament, see Fig. \ref{map2}) is faint in VHE map. The peak in VHE emission is from the north-east region, where the radio emission is faint (see Fig. \ref{map2}).

The azimuthal variation in radio brightness can be either due to the gradient of ambient interstellar medium (ISM) density or due to the gradient of the magnetic field strength at the shock. 
An encounter of SNR G353.6$-$0.7 with a dense molecular cloud can create a density gradient in the ambient ISM, though so far there is no evidence for such an interaction. The $^{12}$CO spectra from the vicinity of SNR G353.6$-$0.7 obtained by \cite{Abramowski2011} does not show any kinematic features that account for a possible interaction of the SNR with a molecular cloud as reported in many interacting SNRs \citep[e.g.] []{moriguchi2005,castelletti2013}. The $\gamma$-ray azimuthal profile of SNR G353.6$-$0.7 is roughly flat \citep{Abramowski2011} except for the two $\gamma$-ray peaks towards the north-east and west, which is suggestive of a fairly uniform ambient ISM density \citep{Abramowski2011}. The shell structure of the SNR recovered in the 325 MHz GMRT map is fairly symmetrical. These three arguments suggest non-interaction of SNR G353.6$-$0.7 with a molecular cloud. Hence it is reasonable to assume that the medium surrounding the SNR is of reasonably uniform density.  

In such a case, the variation in radio brightness can be most likely due to the variation of magnetic field if the injection is isotropic (the efficiency of injection does not depend on the angle between shock normal and magnetic field).   
Radio brightness increases in the region of high magnetic field since synchrotron emissivity is proportional to $B^{3/2}$ for a particle distribution $N(E)$ scaling as $E^{-2}$. If the VHE emission is of leptonic origin, the electrons of energies $E \sim E_{\rm max}$ emit VHE $\gamma$-rays. These electrons experience ample radiative losses in this region and the energy loss rate is proportional to $E^{2}B^{2}$. Thus in the region of high magnetic field, the number of electrons emitting Inverse Compton (IC) $\gamma$-rays deplete faster. This results in a low brightness of VHE emission where the radio brightness is still high \citep{petruk2009a}, thus potentially explain the anti-correlation in the radio and $\gamma$-ray emission. 
This explanation is valid only in the case of synchrotron-loss limited acceleration scenario where $E_{\rm max} \propto$ $B^{-1/2}$ \citep{Reynolds2008}. In the case of age-limited accelaration scenario, $E_{\rm max}$ $\propto B$ \citep{Reynolds2008} and the anti-correlation cannot be explained. However, the age of SNR G353.6$-$0.7 is not well known. The age estimates in the literature range from 2000 to 27000 years \citep{Tian2008,Abramowski2011,fukuda2014,acero2015}.
                                                                                                                                                        
According to this model, we explain the anti-correlated emission of radio and VHE for SNR G353.6-0.7 in a leptonic scenario as follows: Evidence of variation of magnetic field for SNR G353.6-0.7 comes from the analysis of spectral indices in  \S \ref{sec:spec}.  Here we see roughly a 0.5 steepening of the spectral index of filament 2 compared to the typical SNR spectral index of $-$0.5, consistent with synchrotron cooling, indicating higher magnetic field in this filament. Thus the magnetic field strength is relatively high in the region of the south-east and eastern filaments, and hence it is bright in radio. Due to high synchrotron cooling, there are fewer IC electrons, and hence the VHE emission is faint here. The magnetic field strength is low towards the northern region of the SNR, and this results in less energy loss of electrons and bright IC emission. Towards the west, magnetic field strength is likely to be similar or less than that of the northern region, and the synchrotron emission is below the sensitivity limits where VHE emission is bright. One needs to measure local magnetic field using Zeeman splitting of OH maser spots \citep{brogan2000}. Magnetic field can also be estimated from the rim-width of thin X-ray filaments by high-resolution X-ray observations \citep{parizot2006,park2009}. The  roughly a 0.5 steepening of the spectral index of filament 2 coincides with the region where the VHE emission is weakest, supporting the leptonic scenario in which the VHE production is suppressed  due to synchrotron cooling in high magnetic field \citep[][and references therein]{Gabici2016}.  This provides a strong indication of leptonic origin of TeV $\gamma$-rays in SNR G353.6$-$0.7.

 The presence of non-thermal X-ray emission further supports the leptonic origin of VHE emission. Non-thermal synchrotron X-ray emission implies the presence of TeV electrons which can easily produce IC $\gamma$ rays. The absence of any thermal X-ray emission further rules out hadronic origin of VHE emission as thermal X-ray emission is proportional to the square of the gas density, and hadronic processes are likely to dominate in
 high density regions \citep{Gabici2016}. However, the hadronic scenario is still possible if the SNR shock has not been able to heat the gas to high temperature if the gas is clumpy, then the dense clump can produce TeV $\gamma$-rays via hadronic process \citep{Gabici}.

This radio-VHE anti-correlation trend is not seen in any of the other four SNRs in the VHE shell class. In fact, correlated emission of synchrotron radio and IC $\gamma$-ray emission is seen in the bilateral SNR 1006 \citep{petruk2009b,Acero2010}. 
This correlated emission is explained by the variation of maximum energy of electrons over the SNR to compensate for magnetic field variation \citep{petruk2009a}.   
This can also happen if the dependence of injection and the electron maximum energy on the obliquity angle (angle between magnetic field and the normal to the shock) is strong enough to dominate the magnetic field variations \citep{petruk2009a}.

Visual inspection of Figure \ref{map2}, suggests the VHE emission  to be slightly beyond the radio shell. However, the extent of radio emission could be due to sensitivity limitation. An upgraded GMRT observation with 200 MHz bandwidth will provide 2.5 times the sensitivity and excellent $uv$ coverage. Thus it will be more sensitive to the fainter extended features and will be tracing the true extent of the radio emission.

\subsection{Radio counterpart of nearby TeV source HESS J1729$-$345}

There is another TeV source, HESS J1729$-$345,   at position $\alpha_{\rm J2000}$ = 17$^{\rm h}$29$^{\rm m}$35$^{\rm s}$, $\delta_{\rm J2000}$ = $-$34$^{\circ}$32$^{\prime}$22$^{\prime \prime}$,
$\sim$ 30 arcmin away from the centre of HESS J1731$-$347 with no identified counterpart in other wavebands \citep{Abramowski2011}. 
This source is located near a \ion{H}{ii} region (G353.381$-$0.114) and a molecular gas clump in spatial projection \citep{Abramowski2011}. \cite{Abramowski2011} estimate near and far kinematic distances towards this molecular clump at $\sim$ 6 and $\sim$ 10 kpc respectively from $^{12}$CO molecular line observations. If HESS J1729$-$345 is associated with HESS J1731$-$347, which is only 3.2 kpc away, this molecular clump cannot be in the vicinity of HESS J1729$-$345 \citep{Abramowski2011}. 

The TeV emission from HESS J1729$-$345 has been explained by \cite{cui2016} with a hadronic scenario, due to the interaction of the SNR with the surrounding molecular material (as observed from $^{12}$CO emission), via proton--proton collision. In that case, $\gamma$-ray photons would be produced by proton--proton interaction, and synchrotron radio emission is not expected unless the secondary leptons produced by the decay of charged pions radiate significantly. However, in order for this emission to be important, it requires rather high densities and magnetic fields. 

\begin{figure}
	\includegraphics[width=8 cm]{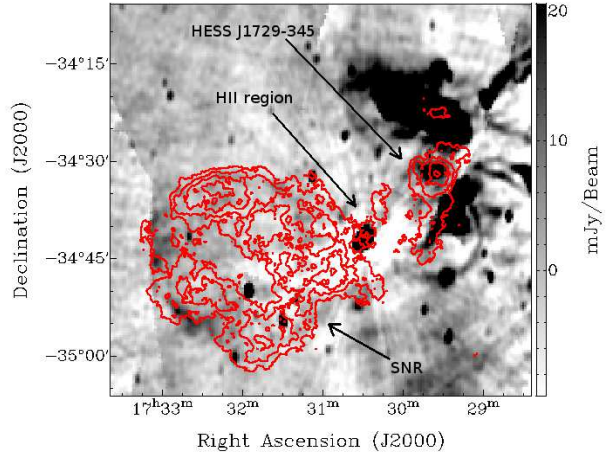}
    \caption{MOST map \citep{green1999} of the field containing G353.6$-$0.7 at 843 MHz overlaid with HESS contours {\bf{with levels 6,8,10,12$\sigma$ ($\sigma$=10 counts)}}. TeV map \citep{Abramowski2011}}.
    \label{byproduct}
\end{figure}

HESS J1729$-$345 was observed in both the 325 and 1390 MHz bands.
Radio emission is not seen in spatial coincidence with HESS J1729$-$345 in the 1390 MHz GMRT map. In the 325 MHz map, the diffuse radio emission towards the west overlaps the region of HESS J1729$-$345 (see Fig. \ref{map2}). The morphology of the emission does not seem to be corresponding to HESS J1729$-$345, which is $\sim$ 7 arcmin gaussian structure. However, a possible radio counterpart of HESS J1729$-$345 is seen in the 843 MHz MOST (see Fig. \ref{byproduct}) and the 1.4 GHz SGPS maps.  The non-detection in the 1390 MHz GMRT map can be due to missing flux. Our GMRT map at 1390 MHz is sensitive to angular scales up to 4 arcmin. Thus significant flux will be missing while imaging a 7 arcmin structure. We obtain a spectral index of 3.0 $\pm$ 0.2, from the 843 MHz MOST map and the 1.4 GHz SGPS map, which is consistent with thermal emission. Thus the radio emission from this possible counterpart is not synchrotron in nature. We note that \cite{cui2016} explained VHE emission from HESS J1729$-$345 in a hadronic scenario. The positive spectral index of the possible radio counterpart supports this explanation.

\section{Summary and conclusions}
We carried out the GMRT observations of HESS J1731$-$347 a.k.a. SNR G353.6$-$0.7 in the  325, 610 and 1390 MHz band. We also observed a nearby source, HESS J1729$-$345 with the GMRT in the 325 and 1390 MHz frequency bands to look for possible radio counterparts. Our conclusions are:
\begin{enumerate}
\item We detect the low-frequency radio morphology of SNR G353.6$-$0.7 with the GMRT at the 325 and 610 MHz frequencies. The shell structure of the SNR with diffuse filaments is revealed in the 325 MHz map, and SNR is partially seen in the 610 MHz map.\\

\item The integrated flux density of the SNR at the 325 MHz frequency is 1.84$\pm$0.15 Jy. Assuming a typical spectral index of $-$0.5 for SNRs, we derive the integrated flux density at the 1.4 GHz as 0.88 Jy. We note that this value is 1.5$\sigma$ lower than the previously reported flux density 2.2$\pm$0.9 Jy at 1.4 GHz \citep{Tian2008}. The change in radio flux density changes the SED slightly.\\

\item The spectral indices of the different filaments are estimated from the 325 and 610 MHz GMRT maps. The values of spectral indices vary from $-$1.11 to $-$0.15 and are consistent with the non-thermal radio emission.\\ 

\item We compare the radio brightness profile with the VHE emission. It is particularly striking that the peak in $\gamma$-ray emission comes from the filament which is faintest in the radio, and the brightest filaments in the radio correspond to faint emission in $\gamma$-rays. The faintest filament in $\gamma$-rays also shows a steep spectral index of $-$1.11$\pm$0.22. This may be due to the effect of non-uniform magnetic field strength, which is suggestive from the possible evidence of synchrotron cooling in that region of the SNR. In this framework, the anti-correlated emission can be explained if the VHE emission is of leptonic origin for an isotropic injection. This kind of an anti-correlated emission is reported for the first time in VHE shell SNRs.\\

\item The visual extent of the VHE emission appears to be slightly beyond the radio shell. However, our radio observations are most likely sensitivity limited. Future radio observations with better sensitivity will probe the true extent of the SNR and the upgraded GMRT observations with 2.5 times higher sensitivity and an order of magnitude larger bandwidth will prove to be useful for this. \\ 

\item We also report a possible radio counterpart of HESS J1729$-$345 from the 843 MHz MOST map and the 1.4 GHz SGPS map. The radio emission is thermal and it supports the hadronic origin of $\gamma$-rays from HESS J1729$-$345. 

\end{enumerate}

\section*{Acknowledgements}
We thank the anonymous referee for constructive inputs. We acknowledge Nissim Kanekar for his support at various stages of this work. P.C. acknowledges support from the Department of Science and Technology via SwaranaJayanti Fellowship award (file no.DST/SJF/PSA-01/2014-15). We thank the staff of the GMRT that made these observations possible. GMRT is run by the National Centre for Radio Astrophysics of the Tata Institute of Fundamental Research.




\begin{thebibliography}{99}
\bibitem[Acero et 
al.(2010)]{Acero2010} Acero, F., Aharonian, F., Akhperjanian, A.~G., et al.\ 2010, \aap, 516, A62 

\bibitem[Acero et al.(2009)]{acero2009} Acero, F., P{\"u}hlhofer, G., Klochkov, D., et al.\ 2009, arXiv:0907.0642 


\bibitem[Acero et 
al.(2015)]{acero2015} Acero, F., Lemoine-Goumard, M., Renaud, M., et al.\ 2015, \aap, 580, A74 


\bibitem[Aharonian et al.(2004)]{Aharonian2004} Aharonian, F.~A., 
Akhperjanian, A.~G., Aye, K.-M., et al.\ 2004, \nat, 432, 75 

\bibitem[Aharonian et al.(2006)]{Aharonian2006} Aharonian, F., Akhperjanian, A.~G., Bazer-Bachi, A.~R., et al.\ 2006, \aap, 449, 223 

\bibitem[Aharonian et al.(2007)]{Aharonian2007} Aharonian, F., 
Akhperjanian, A.~G., Bazer-Bachi, A.~R., et al.\ 2007, \apj, 661, 236 

\bibitem[Aharonian et 
al.(2008)]{Aharonian2008} Aharonian, F., Akhperjanian, A.~G., Barres de Almeida, U., et al.\ 2008, \aap, 477, 353 

\bibitem[Aharonian et al.(2009)]{Aharonian2009} Aharonian, F., 
Akhperjanian, A.~G., de Almeida, U.~B., et al.\ 2009, \apj, 692, 1500 




\bibitem[Bamba et al.(2012)]{bamba2012} Bamba, A., P{\"u}hlhofer, G., Acero, F., et al.\ 2012, \apj, 756, 149 



\bibitem[Basu et al.(2012)]{Basu2012} Basu, A., Mitra, D., 
Wadadekar, Y., \& Ishwara-Chandra, C.~H.\ 2012, \mnras, 419, 1136

\bibitem[Blandford \& Eichler(1987)]{blandford1987} Blandford, R., \& Eichler, D.\ 1987, \physrep, 154, 1 

\bibitem[Blasi(2013)]{blasi2013} Blasi, P.\ 2013, \aapr, 21, 70 

\bibitem[Brogan et al.(2000)]{brogan2000} Brogan, C.~L., Frail, D.~A., Goss, W.~M., \& Troland, T.~H.\ 2000, \apj, 537, 875 



\bibitem[Castelletti et al.(2013)]{castelletti2013} Castelletti, G., Supan, L., Dubner, G., Joshi, B.~C., \& Surnis, M.~P.\ 2013, \aap, 557, L15 


\bibitem[Condon et al.(1998)]{Condon1998} Condon, J.~J., Cotton, 
W.~D., Greisen, E.~W., et al.\ 1998, \aj, 115, 1693 

\bibitem[Cui et al.(2016)]{cui2016} Cui, Y., P{\"u}hlhofer, G., \& Santangelo, A.\ 2016, \aap, 591, A68 

\bibitem[Degrange \& Fontaine(2015)]{degrange2015} Degrange, B., \& Fontaine, G.\ 2015, Comptes Rendus Physique, 16, 587 

\bibitem[Fukuda et al.(2014)]{fukuda2014} Fukuda, T., Yoshiike, S., Sano, H., et al.\ 2014, \apj, 788, 94 


\bibitem[Gabici \& Aharonian(2014)]{Gabici} Gabici, S., \& Aharonian, F.~A.\ 2014, \mnras, 445, L70

\bibitem[Gabici \& Aharonian(2016)]{Gabici2016} Gabici, S., \& Aharonian, F.\ 2016, European Physical Journal Web of Conferences, 121, 04001

\bibitem[Green et al.(1999)]{green1999} Green, A.~J., Cram, 
L.~E., Large, M.~I., \& Ye, T.\ 1999, \apjs, 122, 207 

\bibitem[Green(2011)]{Dave2011} Green, D.~A.\ 2011, Bulletin of the Astronomical Society of India, 39, 289 


\bibitem[Greisen(2003)]{Greisen2003} Greisen, E.~W.\ 2003, in Andr\'e Heck., ed., Information Handling in Astronomy - Historical Vistas. vol. 285, Dordrecht: Kluwer Academic Publishers, 2003., p.109
 
 \bibitem[Halpern \& Gotthelf(2010a)]{halpern2010a} Halpern, J.~P., \& Gotthelf, E.~V.\ 2010a, \apj, 710, 941 

\bibitem[Halpern \& Gotthelf(2010b)]{halpern2010b} Halpern, J.~P., \& Gotthelf, E.~V.\ 2010b, \apj, 725, 1384 

\bibitem[Haslam et 
al.(1982)]{Haslam1982} Haslam, C.~G.~T., Salter, C.~J., Stoffel, H., \& Wilson, W.~E.\ 1982, \aaps, 47, 1 

\bibitem[Haverkorn et al.(2006)]{Haverkorn2006} Haverkorn, M., 
Gaensler, B.~M., McClure-Griffiths, N.~M., Dickey, J.~M., 
\& Green, A.~J.\ 2006, \apjs, 167, 230 

\bibitem[Abramowski et 
al.(2011)]{Abramowski2011} H.E.S.S.~Collaboration, Abramowski, A., Acero, F., et al.\ 2011, \aap, 531, A81 

\bibitem[H.~E.~S.~S.~Collaboration et al.(2016a)]{hess2016a} H.~E.~S.~S.~Collaboration, Abdalla, H., Abdalla, H., et al.\ 2016a, arXiv:1609.08671 

\bibitem[H.~E.~S.~S.~Collaboration et al.(2016b)]{hess2016b} H.~E.~S.~S.~Collaboration, Abdalla, H., Abramowski, A., et al.\ 2016b, arXiv:1611.01863 

\bibitem[H.~E.~S.~S.~Collaboration et al.(2016c)]{hess2016c} H.~E.~S.~S.~Collaboration, Abramowski, A., Aharonian, F., et al.\ 2016c, arXiv:1601.04461 



\bibitem[Jones et al.(1998)]{jones1998} Jones, T.~W., Rudnick, L., Jun, B.-I., et al.\ 1998, \pasp, 110, 125

\bibitem[Landecker 
\& Wielebinski(1970)]{Landecker1970} Landecker, T.~L., \& Wielebinski, R.\ 1970, Australian Journal of Physics Astrophysical Supplement, 16, 1 

\bibitem[Marcote et al.(2015)]{marcote2015} Marcote, B., Rib{\'o}, M., Paredes, J.~M., \& Ishwara-Chandra, C.~H.\ 2015, \mnras, 451, 59 

\bibitem[Moriguchi et al.(2005)]{moriguchi2005} Moriguchi, Y., Tamura, K., Tawara, Y., et al.\ 2005, \apj, 631, 947 

\bibitem[Neugebauer et al.(1984)]{Neugebauer1984} Neugebauer, G., 
Habing, H.~J., van Duinen, R., et al.\ 1984, \apjl, 278, L1 

\bibitem[Orlando et al.(2007)]{orlando2007} Orlando, S., Bocchino, F., Reale, F., Peres, G., \& Petruk, O.\ 2007, \aap, 470, 927 

\bibitem[Parizot et al.(2006)]{parizot2006} Parizot, E., Marcowith, A., Ballet, J., \& Gallant, Y.~A.\ 2006, \aap, 453, 387 

\bibitem[Park et al.(2009)]{park2009} Park, S., Kargaltsev, O., Pavlov, G.~G., et al.\ 2009, \apj, 695, 431 


\bibitem[Petruk et al.(2009a)]{petruk2009a} Petruk, O., Beshley, V., Bocchino, F., \& Orlando, S.\ 2009a, \mnras, 395, 1467 

\bibitem[Petruk et al.(2009b)]{petruk2009b} Petruk, O., Dubner, G., Castelletti, G., et al.\ 2009b, \mnras, 393, 1034 




\bibitem[Reynolds(2008)]{Reynolds2008} Reynolds, S.~P.\ 2008, \araa, 46, 89 

\bibitem[Roger et al.(1999)]{roger1999} Roger, R.~S., Costain, C.~H., Landecker, T.~L., \& Swerdlyk, C.~M.\ 1999, \aaps, 137, 7

\bibitem[Sirothia(2009)]{sirothia2009} Sirothia, S.~K.\ 2009, \mnras, 398, 853

\bibitem[Swarup et al.(1991)]{swarup1991} Swarup, G., Ananthakrishnan, S., Kapahi, V.~K., et al.\ 1991, Current Science, Vol.~60, 95.

\bibitem[Tian et al.(2008)]{Tian2008} Tian, W.~W., Leahy, D.~A., 
Haverkorn, M., \& Jiang, B.\ 2008, \apjl, 679, L85 

\bibitem[Tian et al.(2010)]{Tian2010} Tian, W.~W., Li, Z., 
Leahy, D.~A., et al.\ 2010, \apj, 712, 790 

\bibitem[Wright et al.(2010)]{Wright2010} Wright, E.~L., 
Eisenhardt, P.~R.~M., Mainzer, A.~K., et al.\ 2010, \aj, 140, 1868

\bibitem[Yang et al.(2014)]{yang2014} Yang, R.-Z., Zhang, X., Yuan, Q., \& Liu, S.\ 2014, \aap, 567, A23  
 
\bibitem[Zoonematkermani et al.(1990)]{Zoonematkermani1990} 
Zoonematkermani, S., Helfand, D.~J., Becker, R.~H., White, R.~L., 
\& Perley, R.~A.\ 1990, \apjs, 74, 181 






 


\end{thebibliography}


\label{lastpage}

\end{document}